\documentclass[pdflatex,sn-mathphys-num]{sn-jnl}
\usepackage{graphicx}%
\usepackage{multirow}%
\usepackage{amsmath,amssymb,amsfonts}%
\usepackage{amsthm}%
\usepackage{mathrsfs}%
\usepackage[title]{appendix}%
\usepackage{xcolor}%
\usepackage{textcomp}%
\usepackage{manyfoot}%
\usepackage{booktabs}%
\usepackage{algorithm}%
\usepackage{algorithmicx}%
\usepackage{algpseudocode}%
\usepackage{listings}%

\begin{document}

\title[Article Title]{Response to ``Are Hilbert Spaces Unphysical? Hardly, My Dear!''}

\author*[1]{\fnm{Gabriele} \sur{Carcassi}}\email{carcassi@umich.edu}

\author{\fnm{Robert} \sur{Rozite}}

\author[1]{\fnm{Christine A.} \sur{Aidala}}

\affil[1]{\orgdiv{Physics Department}, \orgname{University of Michigan}, \orgaddress{\street{450 Church St}, \city{Ann Arbor}, \state{MI} \postcode{48109}, \country{United States}}}

%\affil*[2]{\orgdiv{Physics Department}, \orgname{University of Wisconsin-Madison}, \orgaddress{\street{1150 University Ave}, \city{Madison}, \postcode{53706}, \state{WI}, \country{United States}}}

%orcid for Gabriele Carcassi: https://orcid.org/0000-0002-1071-6251

%orcid for Robert Rozite: https://orcid.org/0009-0007-6133-998X

\abstract{A recent criticism of our paper ``The unphysicality of Hilbert spaces'' by Nivaldo Lemos refutes our central argument that a state with finite expectation value can be mapped to a state with infinite expectation value by a coordinate transformation. By conflating coordinate transformation with change of basis in quantum mechanics, Lemos argues that expectation values are invariant under change of variables. In the present work, we clarify the distinction between coordinate transformation and change of basis, and rebut Lemos' main argument.}

\maketitle

%%do we need a section title here?

In our original paper \cite{carcassi}, we claim that Hilbert spaces are problematic because they allow coordinate transformations that map states with finite expectation values to ones with infinite expectations. In his recent criticism \cite{lemos}, Nivaldo Lemos sums up our approach: ``through a change of variables, which is a unitary transformation representing a change of basis, the state $\psi$ in which the expectation value of any positive even power $X^n$ of the position operator is finite has been turned into a state $\phi$ in which these expectation values are infinite.'' He continues to show that expectation values are invariant under changes of basis, and that, therefore, our proposed coordinate transformations cannot change expectation values. The conflation of coordinate transformations and change of basis lies at the core of Lemos' central argument, and will be shown to be inconsistent with established literature. More specifically, Lemos says we reached an ``unsound conclusion'' because we ``interpreted the change of variables as a unitary transformation that changes the state vectors but leaves the observables alone''. This is exactly our procedure, and is consistent with an active transformation as described in various textbooks \cite{gottfried2, ballentine, binney}. From Shankar's ``Principles of Quantum Mechanics'' \cite{shankar}:

\begin{quote}
    Suppose we subject all the vectors \(|V\rangle\) in a space to a unitary transformation
\begin{align}
    |V \rangle \rightarrow U |V \rangle
\end{align}
Under this transformation, the matrix elements of any operator \(\Omega\) are modified as follows: 
\begin{align}
    \langle V' | \Omega | V \rangle \rightarrow \langle UV' | \Omega | UV \rangle = \langle V' | U^\dag \Omega U |V\rangle
\end{align}
It is clear that the same change would be effected if we left the vectors alone and subjected all operators to the change
\begin{align}
    \Omega \rightarrow U^\dag \Omega U
\end{align}
The first case is called an active transformation and the second a passive transformation. The present nomenclature is in reference to the vectors: they are affected in an active transformation and left alone in the passive case. The situation is exactly the opposite from the point of view of the operators. 

Later we will see that the physics in quantum theory lies in the matrix elements of operators, and that active and passive transformations provide us with two equivalent ways of describing the same physical transformation. 

\end{quote}

Coordinate transformations and changes of variables are similar, if not equivalent, concepts that can be interpreted as physical transformations. A simple example is spatial translation, which is carried out as a change of variables: $x \rightarrow x + a$. This example is well established in the literature \cite{gottfried1, binney}, and, intuitively, does not preserve the expectation value of position.

Still, Lemos insists that a coordinate transformation is a unitary transformation representing a change of basis. Given a unitary transformation, \(U\), a change of basis is carried out by transforming states as \(\phi' = U \phi\) and operators as \(O' = UOU^\dag\). Unitarity requires \(UU^\dag = U^\dag U = I\). As such, expectation values are invariant under a change of basis: \(\langle \phi' | O' | \phi' \rangle = \langle \phi | O | \phi \rangle\). Lemos purports that we have failed to transform the position operator in addition to the state, which would be true if we were performing a change of basis. Considering our coordinate transformation as an active transformation, \textit{not} a change of basis, his argument quickly falls apart. Furthermore, he claims ``every time-independent unitary transformation in the Hilbert space of states is a change of basis''. While unitary transformations are used to perform changes of basis, they are more generally used to transform the state of a physical system to another state, without necessarily preserving expectation values. An insistence that states and operators both be transformed by a unitary transformation would preclude us from studying changes of reference frame and time evolution which, through relativity, are intimately linked.

Despite over a hundred years of quantum physics, it is disheartening that a criticism conflating coordinate transformation with change of basis can pass through peer review uncontested. It is our hope that the present work not only rebuts Lemos' central argument, but also clarifies the procedure and intuition behind physical transformations and change of basis in quantum mechanics.

%Lemos further claims that ``any change of basis or change of representation can always be regarded as an active unitary transformation in Hilbert space that changes vectors and operators.'' We find this language to be inconsistent with the literature.

\section*{Acknowledgments}
This paper is part of the ongoing Assumptions of Physics project~\cite{aop-book}, which aims to identify a handful of physical principles from which the basic laws can be rigorously derived. One of the authors of the original paper did not have time to contribute to this response. The other authors thank Robert Rozite for taking the lead in preparing this response.

\section*{Conflict of Interest}
The corresponding author states that there is no conflict of interest.
\section*{Data Availability Statement}
Not applicable.
\bibliography{refs}

\end{document}